\documentclass[useAMS,usenatbib]{mn2e}
\usepackage{graphicx}
\title[Color Variability of SDSSMOC Asteroids]{Color Variability of Asteroids in SDSS Moving Object Catalog}

\author[Gy. M. Szab\'o et al.]{Gy. M. Szab\'o\
$^{1,2}$\thanks{E-mail: szgy@mcse.hu},
\v{Z}. Ivezi\'c$^{3,4}$, M. Juri\'c$^{3}$,
R. Lupton$^{3}$ and L. L. Kiss$^{1,5}$\\
$^{1}$Department of Experimental Physics \& Astronomical Observatory,
University of Szeged,
6720 Szeged, Hungary\\
$^{2}$Department of Physics and Astronomy, Johns Hopkins University,
Baltimore, MD, 21218\\
$^{3}$Princeton University Observatory, Princeton, NJ 08544\\
$^{4}$H.N. Russell Fellow\\
$^{5}$University of Sydney, School of Physics, Sydney, Australia}

\begin{document}

\date{}
\pagerange{\pageref{firstpage}--\pageref{lastpage}} \pubyear{2002}
\maketitle

\begin{abstract}
We report a detection of statistically significant color variations 
for a sample of 7,531 multiply observed asteroids that are listed in 
the Sloan Digital Sky Survey Moving Object Catalog. Using 5-band 
photometric observations accurate to $\sim$0.02 mag, we detect color 
variations in the range 0.06-0.11 mag (rms). These 
variations appear uncorrelated with asteroids physical characteristics 
such as diameter (in the probed 1-10 km range), taxonomic class, and family 
membership. Despite such a lack of correlation, which implies a random nature 
for the cause of color variability, a suite of tests suggest that the detected 
variations are not instrumental effects. In particular, the observed 
color variations are incompatible with photometric errors, and, for objects 
observed  at least four times, the color change in the first pair of observations 
is  correlated with the color change in the second pair. These facts 
strongly suggest that the observed effect is real, and also indicate that for 
some asteroids  color variations are larger than for others. The detected color 
variations can be explained as due to inhomogeneous albedo distribution over an asteroid 
surface. Although relatively small, these variations suggest that fairly large 
patches with different color than their surroundings exist on a significant 
fraction of asteroids. This conclusion is in agreement with spatially resolved 
color images of several large asteroids obtained by NEAR spacecraft and HST. 
\end{abstract}


%\leftline{\Large DRAFT, please do not circulate}

\section{Introduction.}

Asteroids are rotating aspherical reflective bodies which thus exhibit brightness
variations. As recognized long ago (Russell 1906, Metcalf 1907), studies 
of their lightcurves provide important constraints on their physical 
properties, and processes that affect their evolution. For example, well-sampled
and accurate lightcurves can be used to determine asteroid asphericity, 
spin vector, and even the albedo inhomogeneity across the surface (Magnusson
1991). The current knowledge about asteroid rotation rates and lightcurve 
properties is well summarized by Pravec \& Harris (2000). The rotational periods 
range from $\sim$2 hours to $\sim$ 15 hours. The lightcurve amplitudes for 
main-belt asteroids and near-Earth objects are typically of the order 0.1-0.2 mag.
(peak-to-peak). Recently, similar variations have been detected for a dozen Kuiper 
Belt Objects (Sheppard \& Jewitt 2002). The largest amplitudes of $\sim$2 mag. 
(peak-to-peak) are observed for asteroids 1865 Cerberus and 1620 Geographos 
(Wisniewski et al. 1997, Szab\'o et al 2001). 

In contrast to appreciable and easily detectable amplitudes of single-band 
light curves, typical asteroid color variations are much smaller. Indeed,
if albedo didn't vary across an asteroid's surface, then the asteroid 
would not display color variability irrespective of its geometry\footnote{Apart 
from the so-called differential albedo effect (Bowell \& Lumme, 1979).}
While the absence of color variability may also be consistent with a gray 
albedo variation, the strong observed correlation between asteroid albedo 
and color (blue C type asteroids have visual albedo of $p_V\sim0.04$, while
for red S type asteroids $p_V\sim0.15-0.20$, Zellner 1979, Shoemaker et al. 1979) 
implies that non-uniform albedo distribution should be detectable through color 
variability. Following Magnusson (1991), hereafter we will refer to non-uniform albedo
distribution across an asteroid surface as to albedo variegation.

The most notable case of albedo variegation is displayed by 4 Vesta which 
apparently has one bright and one dark hemisphere (Blanco \& Catalano 1979; 
Degewij, Tedesco \& Zellner 1979, Binzel et al. 1997). Definite color variations 
have been detected in only a few dozen asteroids. A color variability at the level 
of a few percent has been measured directly for Eros (V-R and V-I, Wisniewski 1976) 
and for 51 Nemausa (u-b, v-y, Kristensen \& Gammelgaard 1993). In a study that still 
remains one of the largest monitoring programs for color variability, Degewij, 
Tedesco \& Zellner (1979) detected color variations greater than 0.03 mag. in 6 out 
of 24 monitored asteroids. In another notable study, Schober \& Schroll (1982) detected 
color modulation in 49 asteroids. Recently, a spectacular confirmation of albedo 
variegation has been obtained for Eros by NEAR multispectral imaging (Murchie 
et al. 2002). While similar spatially resolved images are available for 
several other objects (e.g. Zellner et al. 1997,  Binzel et al. 1997, Baliunas et al. 
2003), the number of asteroids with observational constraints on their albedo 
variegation remains small.

Here we study asteroid color variability by utilizing the Sloan Digital Sky 
Survey Moving Object Catalog (hereafter SDSSMOC, Ivezi\'{c} et al. 2002a).
SDSSMOC currently contains accurate (0.02 mag) 5-band photometric measurements 
for over 130,000 asteroids. A fraction of these objects are previously 
recognized asteroids with available orbits, and 7,531 of them were observed 
by SDSS at least twice. We use the color differences between the two observations
of the same objects to constrain the ensemble properties, as opposed to studying 
well-sampled light curves for a small number of objects. The lack of detailed information
for individual objects is substituted by the large sample size which allows us to 
study correlations between color variability and various physical properties in a 
statistical sense. Also, objects in the sample studied here have typical sizes 
1--10 km, about a factor 10 smaller than objects for which color variations
have been reported in the literature.

We describe the SDSSMOC and data selection in Section 2, and in Section 3 we 
perform various tests to demonstrate that detected color variability of multiply 
observed objects is not an observational artefact. In Section 4 we search for 
correlations between the color variability and asteroid physical properties, and 
summarize our results in Section 5.

\section{           SDSS Observations of Moving Objects      }

SDSS is a digital photometric and spectroscopic survey which will cover 10,000 deg$^2$ 
of the Celestial Sphere in the North Galactic cap and a smaller ($\sim$ 
225 deg$^2$) and deeper survey in the Southern Galactic hemisphere (Azebajian 
et al. 2003, and references therein). The survey sky coverage will result in photometric 
measurements for about 50 million stars and a similar number of galaxies. About 50\% 
of the Survey is currently finished. The flux densities of detected objects are 
measured almost simultaneously in five bands (Fukugita et al. 1996; $u$, $g$, $r$, $i$, 
and $z$)  with effective wavelengths of 3551 \AA, 4686 \AA, 6166 \AA, 7480 \AA, and 8932 \AA, 
95\% complete for point sources to limiting magnitudes of 22.0, 22.2, 22.2, 21.3, and 
20.5 in the North Galactic cap. Astrometric positions are accurate to about 
0.1 arcsec per coordinate (rms) for sources brighter than 20.5$^m$ (Pier et al. 2002), 
and the morphological information from the images allows robust star-galaxy separation 
(Lupton et al. 2001) to $\sim$ 21.5$^m$.

SDSS, although primarily designed for observations of extragalactic objects, is
significantly contributing to studies of the solar system objects, because asteroids
in the imaging survey must be explicitly detected to avoid contamination of the
samples of extragalactic objects selected for spectroscopy. Preliminary analysis
of SDSS commissioning data (Ivezi\'{c} et al. 2001, hereafter I01) showed that SDSS 
will increase the number  of asteroids with accurate five-color photometry by more 
than two orders of magnitude, and to a limit about five magnitudes fainter (seven 
magnitudes when the completeness limits are compared) than previous multi-color 
surveys (e.g. The Eight Color Asteroid Survey, Zellner, Tholen \& Tedesco 1985).

\subsection{ The Sample Selection Using SDSS Moving Object Catalog }

SDSS Moving Object Catalog\footnote{Available at http://www.sdss.org} is a public,
value-added catalog of SDSS asteroid observations. In addition to providing SDSS 
astrometric and photometric measurements, all observations are matched to known 
objects listed in the ASTORB file (Bowell 2001), and to the database of proper orbital 
elements (Milani et al. 1999), as described in detail by Juri\'{c} et al. (2002, 
hereafter J02). Multiple SDSS observations of objects with known orbital parameters can be accurately 
linked, and thus SDSSMOC contains rich information about asteroid color variability. 

We select 7,531 multiply observed objects from the second SDSSMOC edition (ADR2.dat)
by requiring that the number of observations ($N_{ap}$) is at least two, and use the 
first and second observations to compute photometric changes. A ``high-quality'' sample 
of 2,289 asteroids is defined by two additional restrictions: 
\begin{itemize}
\item
In order to avoid the increased photometric errors at the faint end, we 
require $r < 19$.
\item
In order to minimize the effect of variable angle from the opposition on the observed 
color, we select only objects for which the change of this angle is less than 1.5 degree. 
\end{itemize}

We also utilize a subsample of 541 asteroids that were observed at least four times.

Since SDSS observations of asteroids are essentially random, and the time between
them (days to months, 75\% of repeated observations are obtained within 3 months,
and 86\% within a year) is much longer than typical rotational periods ($<$ 1 day),
the phases of any two repeated observations are practically uncorrelated. Thus, 
the distribution of magnitude and color changes for a large ensemble of asteroids 
is a good proxy for random two-epoch sampling of an asteroid's single-band and color 
lightcurves. Of course, this is strictly true only if the distributions of amplitudes and 
shapes of these light curves are fairly narrow. For wide distributions of amplitudes
and shapes, the observed two-epoch magnitude and color changes represent convolution
of the two effects. 

%Fig. 1. 
\label{Fig1}
\begin{figure} 
\centering
\includegraphics[bb=40 250 580 750, width=\columnwidth]{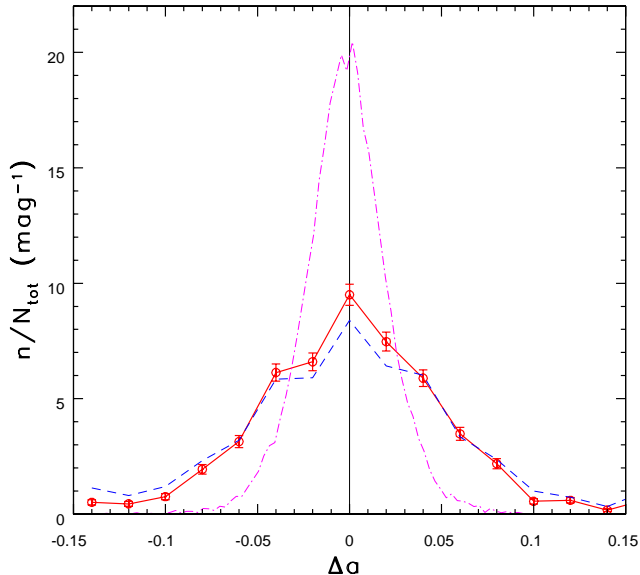}
\caption{
The dashed line shows the distribution of the $a$ color change between two 
epochs for all 7,531 asteroids from SDSSMOC that were observed at least twice. 
The symbols with error bars connected by the solid line show the distribution 
of the $a$ color change for a subset of 2,289 asteroids brighter than $r=19$ and
with the difference in angles from the opposition smaller than 1.5 degree. 
Its equivalent Gaussian width, determined from the interquartile
range, is 0.053 mag. The dash-dotted line shows the distribution of
the $a$ color change between two epochs for 21,000 stars brighter
than $r=19$. Its width, which indicates the measurement error for 
$a$ color, is 0.023 mag.}
\end{figure}

\section{      Asteroid Color Variability in SDSSMOC  }

The colors of asteroids in SDSS photometric system are discussed in detail by I01.
They defined a principal color in the $r-i$ vs. $g-r$ color-color diagram, $a$, as
\begin{equation}
            a \equiv 0.89 \,(g - r) + 0.45 \,(r - i) - 0.57.
\end{equation}

The $a$ color distribution is strongly bimodal (see Fig.~9 in I01), with the 
two modes at $-0.1$ and $0.1$. The rms scatter around each mode is about 0.05 mag.
The two modes are associated with different taxonomic classes: the ``blue'' 
mode includes C, E, M and P types, and the ``red'' mode includes S, D, A, V and J
types (see Fig. 10 in I01). The Vesta type asteroids (type V) can be effectively
separated from ``red'' asteroids using the $i-z$ color (J02). The $a$ and $i-z$ 
colors are strongly correlated with dynamical family membership (Ivezi\'{c} et al. 
2002c, hereafter I02c). Hereafter, we chose the $a$ color as the primary 
quantity to study asteroid color variability.

%Fig. 2. 
\label{Fig2}
\begin{figure} 
\centering
\includegraphics[bb=20 20 501 822, width=\columnwidth]{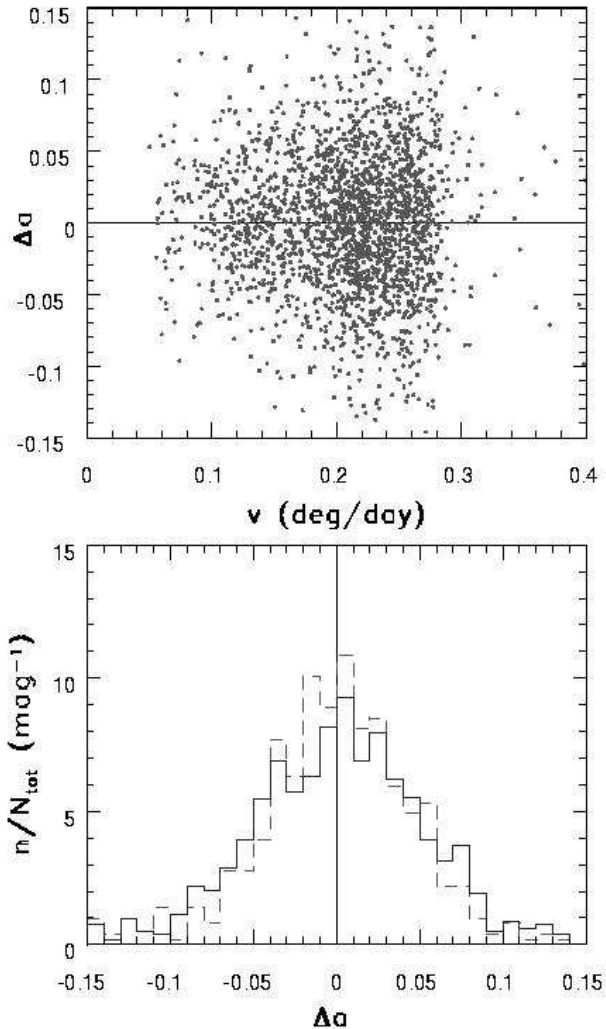} 
\caption{
The upper panel shows the $a$ color change between two epochs
for the same asteroids as in in Fig. 1, as a function of the object's
velocity.  The bottom panel compares the width of the distribution
of the $a$ color change for 511 objects with $0.10 < v < 0.18$
(dashed line) and for 1,065 objects with $v > 0.22$ deg/day (solid
line). Note that the two histograms are statistically indistinguishable
indicating that the color measurement is not affected by the object's
apparent velocity.}
\end{figure}

%Fig. 3. 
\label{Fig3}
\begin{figure} 
\centering
\includegraphics[bb=20 20 501 822, width=\columnwidth]{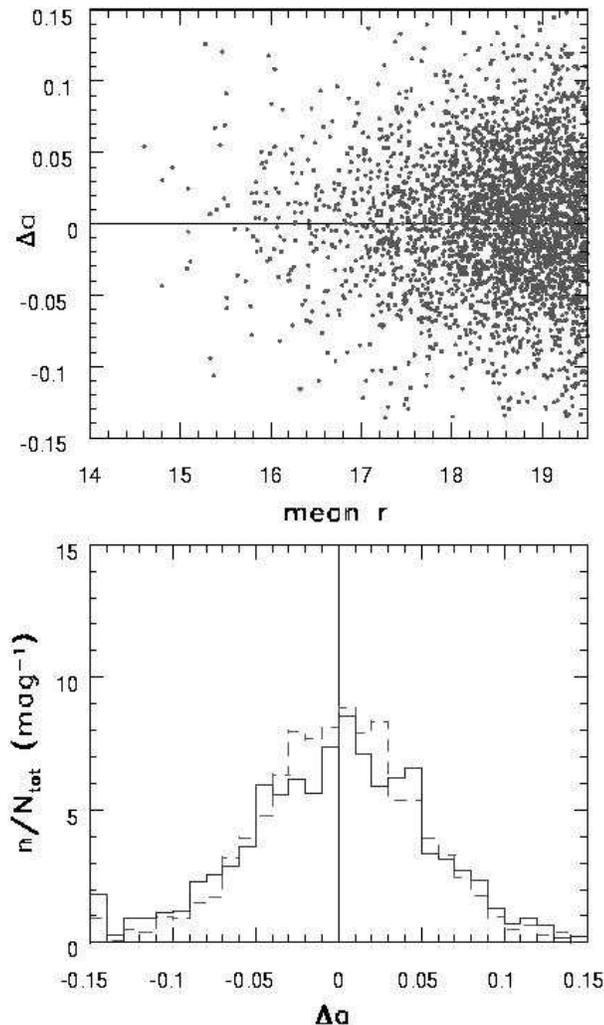} 
\caption{
The upper panel shows the $a$ color change between two epochs
for the same asteroids as in in Fig. 1, as a function of the object's
apparent magnitude.  The bottom panel compares the width of the
distribution of the $a$ color change for 214 objects with $r < 17$
(dashed line) and for 898 objects with $18.5 < r < 19$ (solid line).
Note that the two histograms are statistically indistinguishable
indicating that the color measurement is not affected by the object's
apparent magnitude.}
\end{figure}

%Fig. 4. 
\label{Fig4}
\begin{figure} 
\centering
\includegraphics[bb=20 20 501 822, width=\columnwidth]{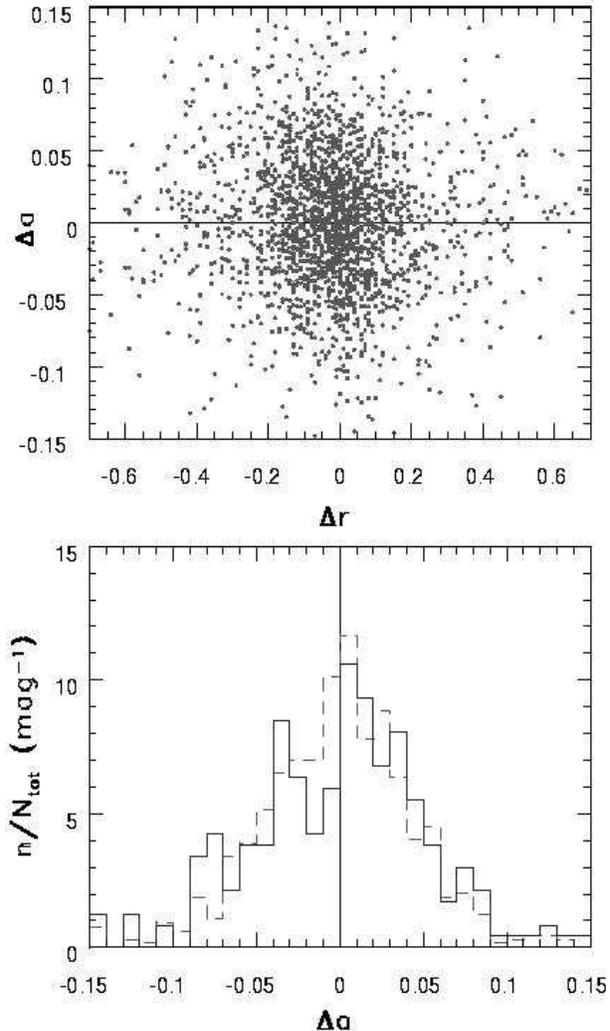} 
\caption{
The upper panel shows the $a$ color change between two epochs
for the same asteroids as in in Fig. 1, as a function of the
change in the $r$ band apparent magnitude  The bottom panel compares
the width of the distribution of the $a$ color change for 645 objects
with $|\Delta(r)| < 0.05$ (dashed line) and for 1,644 objects with
$|\Delta(r)| > 0.05$ (solid line). Note that the two histograms are
statistically indistinguishable indicating that the color measurement
is not correlated with the change of the object's apparent magnitude.}
\end{figure}

%Fig. 5. 
\label{Fig5}
\begin{figure} 
\centering
\includegraphics[bb=20 20 501 822, width=\columnwidth]{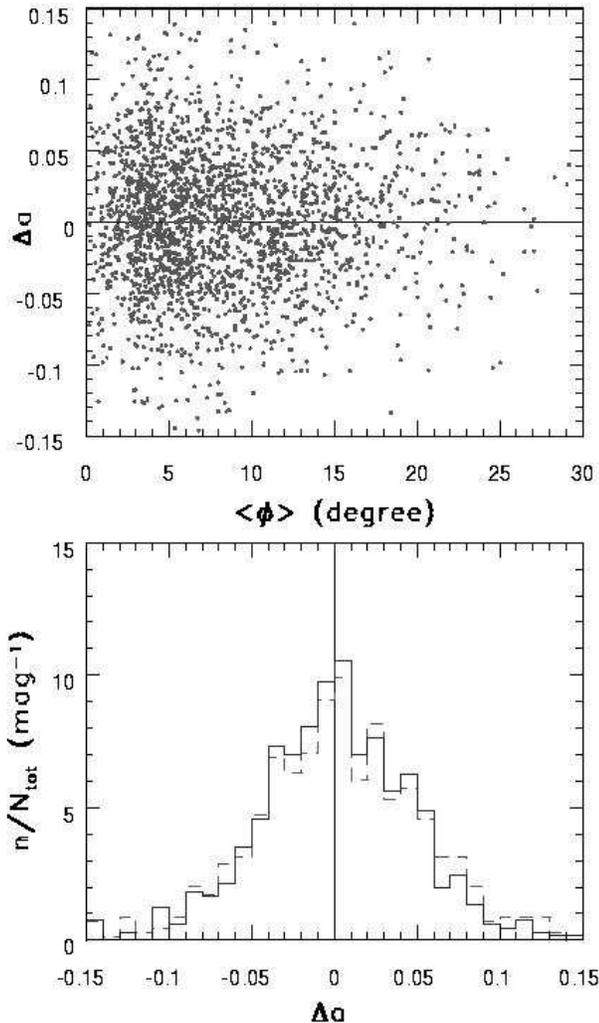} 
\caption{
The upper panel shows the $a$ color change between two epochs
for the same asteroids as in in Fig. 1, as a function of the mean
angle from the opposition. The bottom panel compares the width of the 
distribution of the $a$ color change for 712 objects with $\phi < 5^\circ$ 
(dashed line) and for 664 objects with $10^\circ < \phi < 20^\circ$ (solid line). 
Note that the two histograms are statistically indistinguishable indicating 
that the color measurement is not correlated with the mean angle from the opposition.}
\end{figure}

%Fig. 6
\label{Fig6}
\begin{figure} 
\centering
\includegraphics[bb=20 20 501 822, width=\columnwidth]{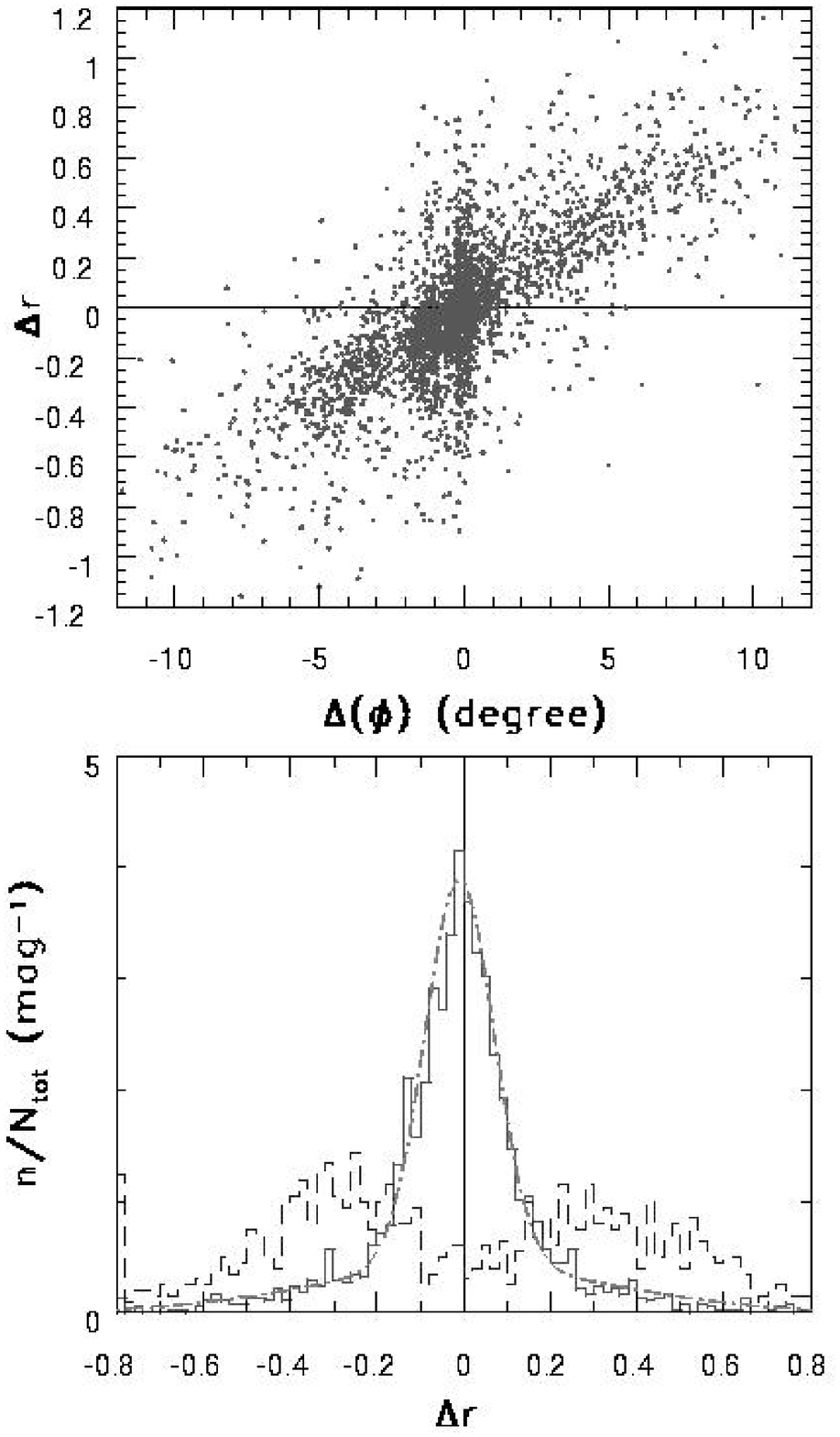} 
\caption{
The upper panel shows the change of $r$ magnitude for the full sample,
as a function of the change of the angle from the opposition. The bottom panel 
compares the $\Delta r$ distributions for two ranges of the angle change: 
$0^\circ - 1^\circ$ as the solid line and $2^\circ - 10^\circ$ as the 
dashed line. The former is well fit by a sum of two Gaussians (see text)
shown as the dash-dotted line.}
\end{figure}

\subsection{  Detection of Asteroid Color Variability in SDSSMOC  }

The observed distribution of the $a$ color change between two epochs,
$\Delta a$, for 2,289 selected asteroids, is shown by symbols (with error
bars) in Figure 1. Its equivalent Gaussian width determined from the interquartile
range (hereafter ``width''), is 0.053 mag. The expected width based on the
formal errors reported by the SDSS photometric pipeline (``photo'', Lupton
et al. 2001) is 0.02 mag., indicating that the observed $\Delta a$
distribution reflects intrinsic asteroid color changes. However, the formal
errors may not be correct.  In order to determine the measurement accuracy
for the $a$ color, we use 21,000 stars with $r < 19$ that were observed
twice. The dashed line in Figure 1. shows their $\Delta a$ distribution. 
Its width is 0.023 mag., in agreement with the expectations based on formal errors.

Subtracting the error distribution width of 0.023 mag. in quadrature, the 
intrinsic $a$ color root-mean-square (rms) variation is 0.04 mag. While Fig.1 
demonstrates that this measurement is statistically highly significant, in the 
remainder of this section we test for the presence of spurious observational
effects that could be responsible for the observed asteroid color variation.

\subsection{ Tests for Spurious Observational Effects }

\subsubsection{    Color Change vs. Asteroid's Velocity }

Although the formal photometric errors for stars are correct, asteroids move 
during observations and their motion could in principle affect the photometric
accuracy. While this effect should be negligible (the images are not strongly trailed), 
we test for it by correlating $\Delta a$ with the object's velocity. If the photometric accuracy
is lower for moving objects, the  $\Delta a$ distribution width should increase 
with magnitude of the object's velocity. The $\Delta a$ vs. $v$ diagram 
is shown in the top panel in Figure 2. The bottom panel compares the width 
of the $\Delta a$ distribution for two subsamples selected by velocity.
The dashed line show the $\Delta a$ distribution for 511 objects with 0.1
deg/day $ < v <$ 0.18 deg/day, and the solid line for 1,065 ones with $v>$0.22 
deg/day. As evident, the two histograms are statistically indistinguishable.

\subsubsection{    Color Change vs. Apparent Brightness   }

The photometric errors typically increase with apparent magnitude of the measured
object. While the adopted faint limit ($r<19$) is sufficiently bright that SDSS photometric
errors are nearly independent of magnitude (Ivezi\'{c} et al. 2002c), 
we test this expectation by correlating $\Delta a$ with the mean apparent magnitude
in Fig. 3. The bottom panel compares the width of the $\Delta a$ distribution 
for two subsamples selected by $r < 17$ (dashed line, 214 objects) and by 
$18.5 < r < 19$  (solid line, 898 objects). There is no significant difference 
between the two histograms. Relaxing the magnitude cut to $r<20$ yields the
same null result with an increased sample size by a factor of 2. We have also
tested $\Delta a$  vs. $\Delta r$ dependence, illustrated in Fig. 4, and did not 
detect any correlation.

\subsubsection{    Color Change vs. the Angle from the Opposition  }

In order to minimize the opposition effect (see Section 6.1 in
I01), the maximum change of the angle from the opposition (between two 
observations) in the analyzed sample is constrained to 1.5 deg (this
limit may be slightly too conservative since we find no correlation 
between $\Delta a$ and the {\em change} of the angle from the opposition
for changes as large as 10$^\circ$). To exclude the possibility that 
the color change is affected by the {\em mean} angle from the opposition,
we analyze $\Delta a$ as a function of this angle in Fig. 5. 
There is no discernible correlation. 

Unlike the color change which is not correlated with the angle from
the opposition, the change of magnitudes between two epochs is strongly
correlated with this angle, as expected. The top panel in Fig. 6 shows the 
dependence of $\Delta r$ on the change of the angle from the opposition.
The comparison of $\Delta r$ distributions for two ranges of the angle
change ($0^\circ - 1^\circ$ as the solid line and $2^\circ - 10^\circ$ as the 
dashed line) are shown in the bottom panel. As evident, even when asteroids are
observed at practically the same position, $\Delta r$ has a wide distribution, 
mostly due to asteroid rotation. We find that the $\Delta r$ distribution can be
well fit by a sum of two Gaussians with the widths of 0.08 mag and 0.35 mag
and amplitude ratio 2:1, shown as the dash-dotted line. The measured 
$\Delta r$ distribution can be used to constrain the distributions of 
asteroid axes ratios and rotational periods. Such an analysis, while 
interesting in its own right, is beyond the scope of this paper.

\subsubsection{The Effects of Non-simultaneous Measurements}
\label{nsm}

%Fig. 7. 
\label{Fig7}
\begin{figure*}
\centering
\includegraphics[bb=20 20 575 427, width=2\columnwidth]{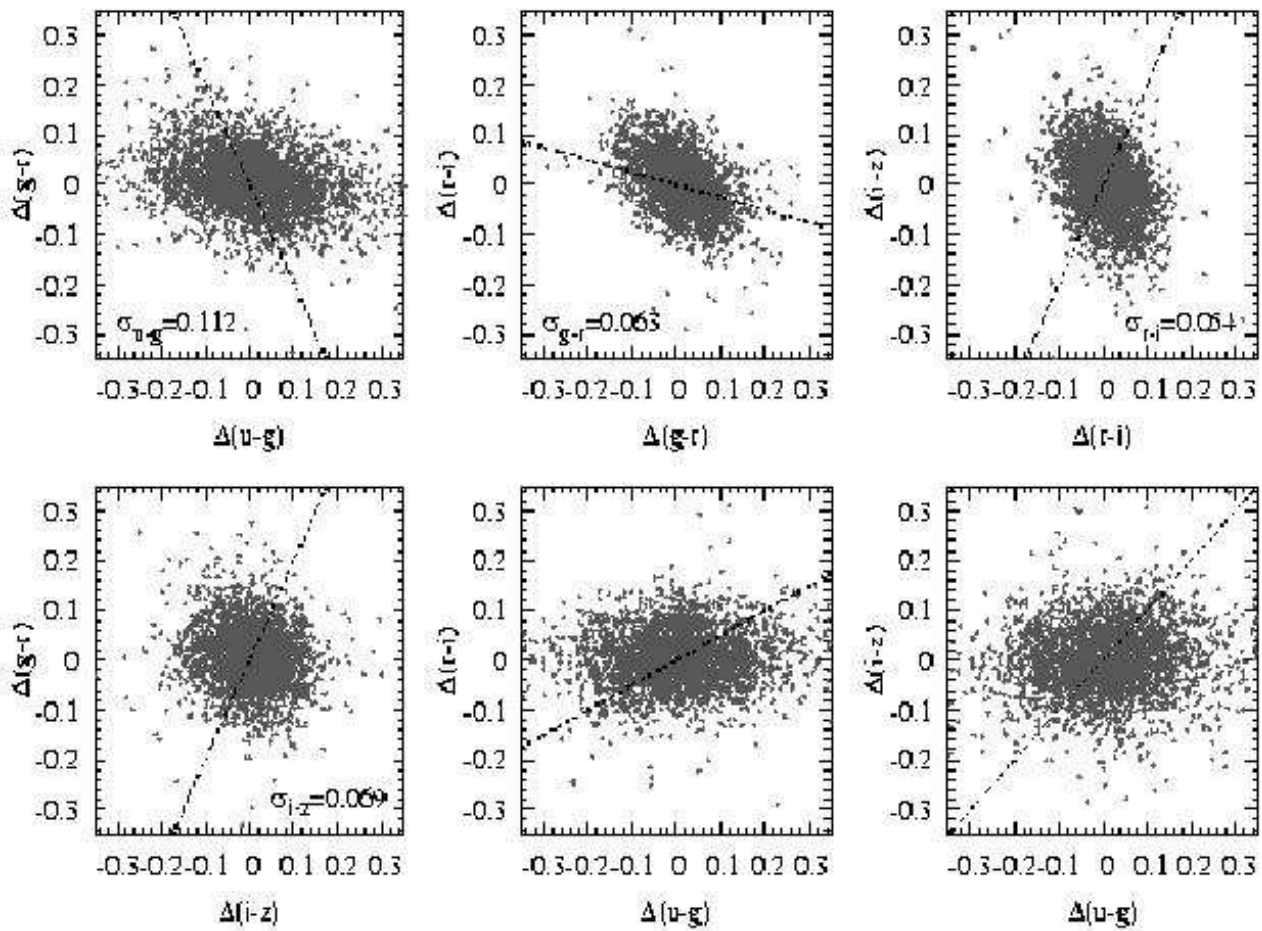} 
\caption{Correlations between the changes of four SDSS colors for the same 
asteroids as in Fig. 1. The dashed lines show the expected slopes 
if non-simultaneous observations and fast, large-amplitude
variability produce significant bias in color measurements (see \S3.2.4).
This bias is not supported by the displayed data.}
\end{figure*}

The SDSS photometric measurements are obtained in the order r-i-u-z-g,
and the elapsed time between the first (r) and last (g) measurement
is $\sim5$ minutes. The asteroid brightness variation during this time,
even if achromatic, introduces a bias in the color measurement. For example,
if an asteroid is observed during the rising part of its lightcurve, 
the $r-i$ color is biased red, and the $g-r$ color is biased blue. Also, 
due to fixed filter order, the color biases should be correlated. The 
expected correlations are $\Delta(g-r)/\Delta(u-g) \sim -2$, 
$\Delta(r-i)/\Delta(g-r) \sim -0.25$, $\Delta(i-z)/\Delta(r-i) \sim 2$,
$\Delta(g-r)/\Delta(i-z) \sim -2$, 
$\Delta(r-i)/\Delta(u-g) \sim 0.5$, $\Delta(i-z)/\Delta(u-g) \sim 1$.
With conservative assumptions that typical rotational period is as short as 
2 hours, and that the peak-to-peak amplitude is as large as 0\fm5, 
the expected rms contribution is the largest for $g-r$ color and 
equal to $\sim$0\fm03, while for $r-i$ color it is less than 0\fm01.

Fig. 7 shows the six possible correlations of the four SDSS colors. The rms
scatter of each color is also displayed in the figure. As evident, the 
variations in all four colors are too large by a factor of few to be explained 
by any plausible rotation parameters. Furthermore, the slopes of expected 
correlations, shown by the dashed lines, are not supported by the data. We 
conclude that the non-simultaneous nature of SDSS color measurements is not
significantly contributing to the observed color variation.  

The $u-g$ and $i-z$ color changes show the largest width. We have verified
that imposing a magnitude cut ($<19$) on the $u$ and $z$ magnitudes does not 
decrease the scatter, i.e. the large widths are not caused by the lower 
sensitivity of these two bands.  

\subsubsection{         Correlated color variations}

Figure 7 demonstrates that some color pairs (e.g. $\Delta(r-i)$ vs. 
$\Delta(g-r)$ show correlations with a slope which cannot be explained by 
the examined instrumental effects. We characterize the observed correlations 
by their linear regression, the $rms$ of the fit and the linear correlation coefficient $C$. 
All the fits cross the origin (the 0th-order coefficients do not differ 
significantly from 0). We determined the following correlations:

\begin{eqnarray}
\Delta(g-r)=-0.13(1)\cdot \Delta(u-g) \\ 
\nonumber \ \ \ rms=0.06 \ \ \ C=-0.26\\
\Delta(r-i)=-0.40(2)\cdot \Delta(g-r) \\ 
\nonumber \ \ \ rms=0.06 \ \ \ \ C=-0.42\\
\Delta(i-z)=-0.51(2)\cdot \Delta(r-i) \\ 
\nonumber \ \ \ rms=0.07 \ \ \ \ C=-0.42\\
\Delta(i-z)=-0.13(2)\cdot \Delta(g-r) \\ 
\nonumber \ \ \ rms=0.07 \ \ \ \ C=-0.12
\end{eqnarray}

\noindent 
There are some noteworthy aspects of this test. First, the change of
$u-g$ color does not correlate with the change of $r-i$ and $i-z$ colors, 
although the scatter of $\Delta(u-g)$ is much higher than for the other colors. 
Also, the color indices $u-X$, where $X=g, r, i, z$, do not correlate with 
any other color index that does not include $X$. This is consistent with
a hypothesis that the blue part of the spectrum is affected by a process
which causes no color variation of the red colors.

\subsubsection{ The Variability Induced Motion in Color-color Diagrams }

The changes of $r-i$ and $g-r$ colors, which are used to define color $a$, 
seem to be weakly correlated, with the median slope $\Delta(r-i) \sim 
-0.4\,\Delta(g-r)$ (3). This implies that, on average, 
the line connecting two observations in this diagram is more aligned with 
the second principal axis which is perpendicular to $a$, than with the $a$ 
axis. Using the definition of the second principal color, hereafter named $p$, 
\begin{equation}
         p \equiv 0.45 \,(g - r) - 0.89 \,(r - i) - 0.11,
\end{equation}
we constructed the principal color diagram shown in the top panel in Fig. 8.
Note that this diagram is simply a rotated version of the $r-i$ vs. $g-r$
diagram. In this panel we simply plot the mean value of each principal 
color, while in the middle panel we connect two individual measurements
by lines, for a small subset of objects with $18 < r < 18.3$ (to avoid crowding)
and a change in each color of at least 0.03 mag. As evident, for objects with large 
color variations, the changes of the two principal colors seem to be
somewhat correlated, with the slope larger than 1. That is, the $p$ color 
varies more than the $a$ color. In the bottom panel we show the 
change of $p$  color vs. the change of $a$ color for all the objects in the 
``high-quality'' subsample, as well as the rms scatter in each color. 

The observed variability of the $p$ color appears nearly sufficient to explain 
its single-measurement distribution width of $\sim 0.05$ mag. (the rms width of 
the change in that color is 1.33 of its single-epoch width, i.e. only slightly 
smaller than the expected value of 1.41 if the variability was the only reason 
for its finite value). In order to test this hypothesis, we use a subsample of 
541 asteroids that were observed at least four times. The distribution width 
for the mean $p$ color, obtained by averaging the four measurements, is expected
to be smaller than that measured for any individual epoch. On the other hand,
no significant difference in the distribution widths is expected for the 
$a$ color. The distributions shown in Fig. 9 suggest that the variability 
contributes significantly to the $p$ color distribution width, and much
less to the width of each individual mode in the $a$ color distribution.

Since the variability induced motion in the $p$ vs. $a$ color-color diagram 
is more aligned with the $p$ axis than with the $a$ axis,  over 90\% of asteroids 
have the same $a$-based classification ($<0$ vs. $>0$) in different epochs. 
The same behavior also indicates that the color variability cannot be explained 
as due to mixing of the two basic materials, corresponding to C and S types, 
on the asteroid surfaces. We will return to this point in Section 5.

%Fig. 8. 
\begin{figure} 
\label{Fig8}
\centering
\includegraphics[bb=10 20 382 812, width=\columnwidth]{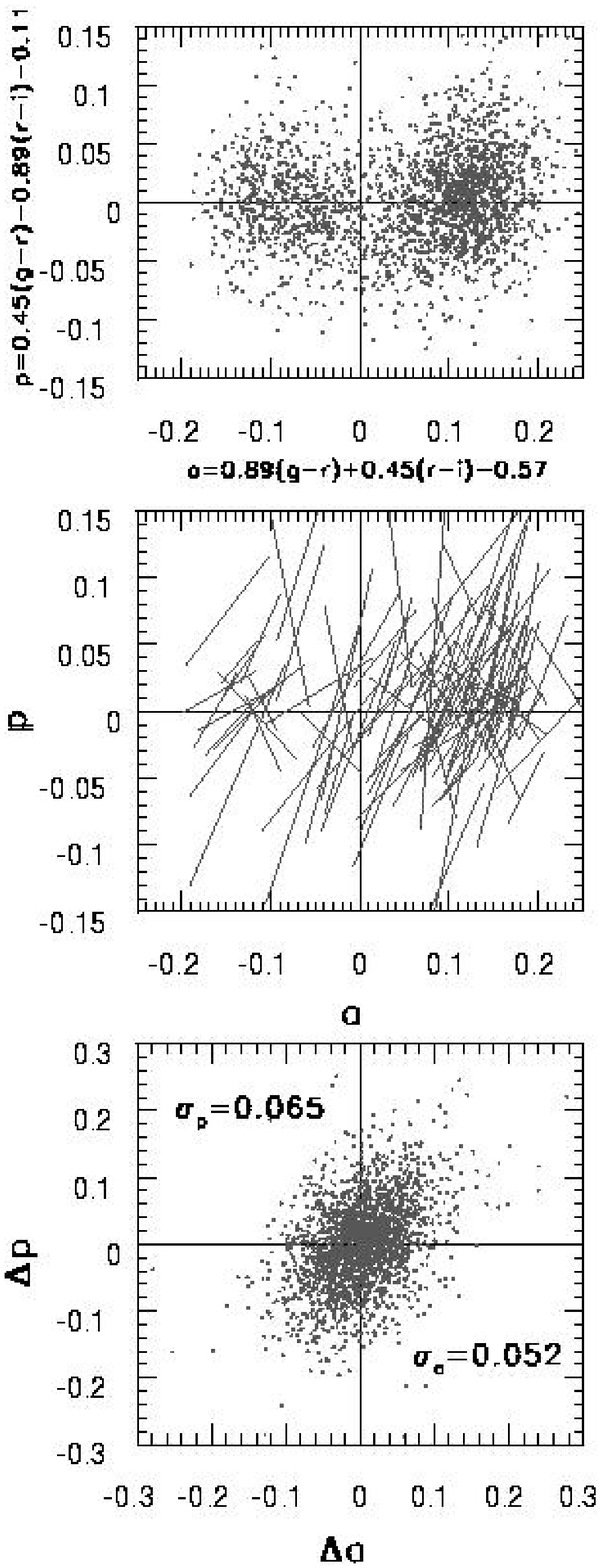} 
\caption{
The top panel shows the asteroid principal color diagram, constructed
with the mean colors for two measurements. In the middle panel 
two individual measurements are connected by lines, for a small subset of 
objects with $18 < r < 18.3$ and a change in each color of at least 0.03 mag.
Note that for objects with large color variations, the changes of the two 
principal color seem to be somewhat correlated. The bottom panel show the 
change of $p$ color vs. the change of $a$ color for all the objects in 
the ``high-quality'' subsample, as well as the rms scatter in each color.
}
\end{figure}

%Fig. 9. 
\label{Fig9}
\begin{figure} 
\centering
\includegraphics[bb=140 100 520 735, width=\columnwidth]{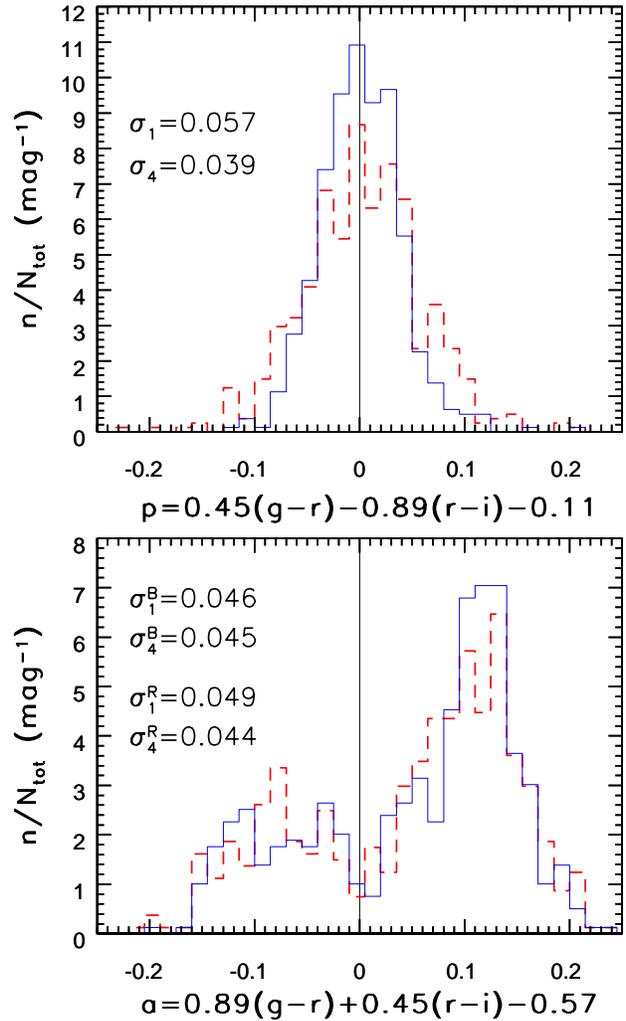} 
\caption{The comparison of the $p$ (top panel) and $a$ (bottom panel) color 
distributions obtained by averaging four measurements (solid lines) for 541
asteroids, and those for individual measurements (dashed lines). The $\sigma$
values are the distribution widths, subscripted 1 for single epoch and 4 for 
average measurements. The widths for the $a$ color distribution are determined
separately for objects with $a<0$ (superscripted B) and $a>0$ (R). Note that
the width of the $p$ color distribution is 1.46 larger for single epoch
measurements, while the $a$ color distribution does not change appreciably.
}
\end{figure}

\subsection{    The   Repeatability  of  Color   Variations    }

The suite of tests in this Section suggest that the observed color variations are 
not an artefact (either observational, or caused by phenomena such as differential 
opposition effect). However the lack of any correlation with physical parameters,
such as color, size and family membership, discussed in the next Section, 
may be elegantly explained as caused by some ``hidden'' random error contribution, 
for example a problem introduced by the processing software. Here we present a test 
which demonstrates that at least in one aspect the observed color variation is 
not random. 

If it is true that asteroids exhibit a varying degree of color variability,
as they do, for example, for single-band variability, then the color changes 
detected in two independent pairs of observations may correlate to some degree.
At the same time, such a correlation would give credence to the measurements
reliability. We use a subsample of 541 asteroids that were observed at least four 
times to test whether such a correlation exist. The top panel in Fig. 10 
plots the change of $a$ color in a pair of observations vs. the change in another
independent pair of observations. We have verified that the marginal distributions 
in each coordinate are indistinguishable. 

To illustrate this point, in the bottom panel we compare the distributions of the 
$a$ color change in one pair of observations for two subsamples selected by the $a$ 
color change in the other independent pair observations, as marked by the dashed
lines in the top panel. If the color changes in two pairs of observations are 
uncorrelated, then the two histograms should be indistinguishable. However, they are 
clearly different, indicating that these independent observations ``know'' about 
each other!

%Fig. 10. 
\label{Fig10}
\begin{figure} 
\centering
\includegraphics[bb=140 100 520 735, width=\columnwidth]{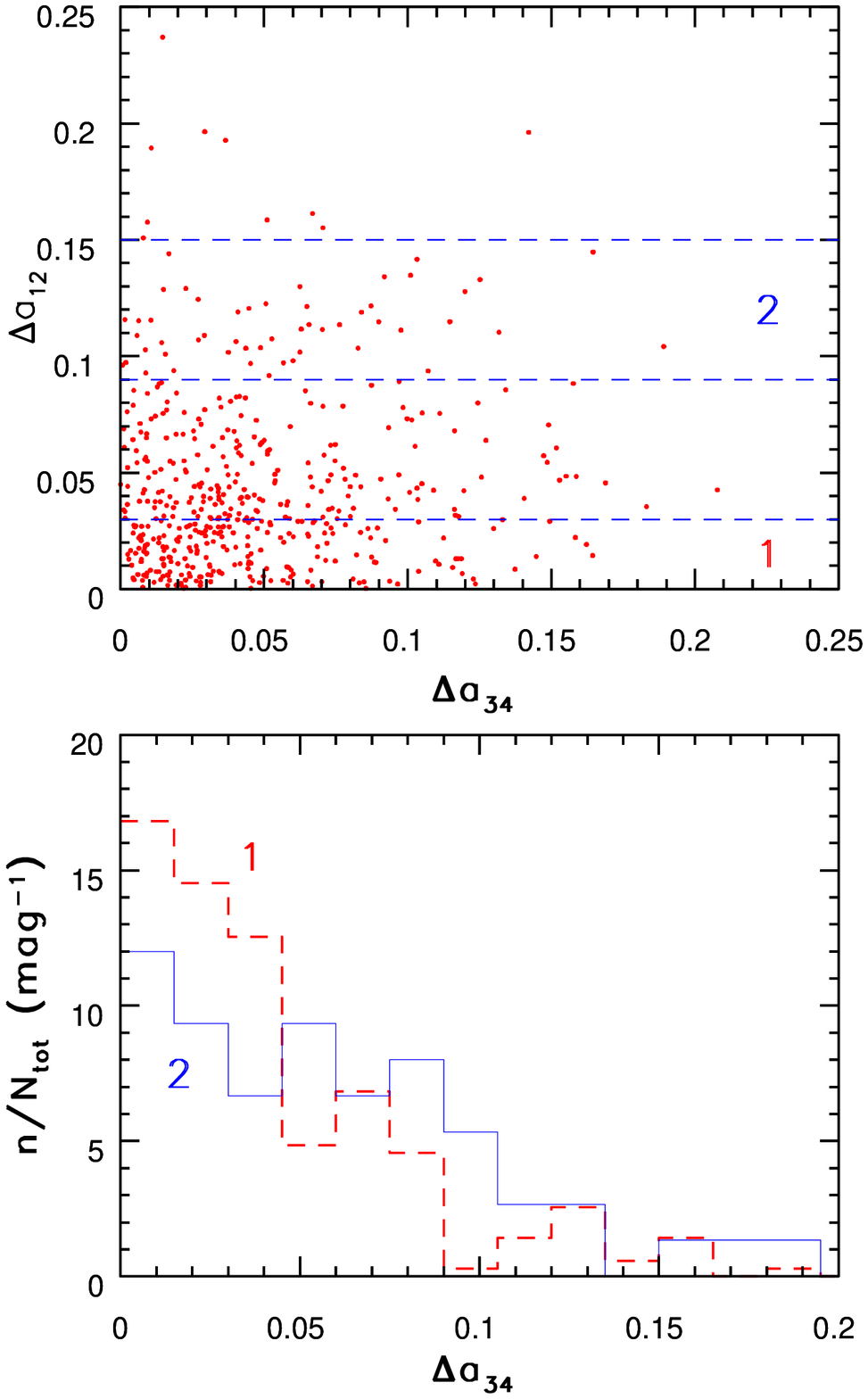} 
\caption{
The top panel plots the changes of $a$ color in two independent pairs of 
observations (change in the first pair vs. the change in the second pair)
for 541 asteroids observed at least four times. The bottom panel compares 
the histograms of $a$ color change in one of the two pairs of observations, 
for two subsets selected by the $a$ color change in the other pair of 
observations, as marked by the dashed lines in the top panel (see text). 
Note that the two distributions are different, indicating that these independent 
observations ``know'' about each other. The same conclusion is obtained when
the axes are reversed.}
\end{figure}

In order to quantify the statistical significance of the difference  between
the two histogram, we perform two tests. A two-sample Kolmogorov-Smirnov 
test (Lupton 1993) indicates that the histograms are different
at a confidence level of more than 99\%. Another test, proposed by 
Efron and Petrosian (1992), which uses the entire 2-D sample, indicates that
the two variables are correlated at a confidence level of 95\%. 
The somewhat lower confidence level than for the first test is probably
due to the contribution of points with small color changes, whose distribution 
may be randomized by photometric errors.
 
We conclude that objects with large color variations in one pair
of observations tend to show relatively large color variation in the other,
independent, pair of observations. This difference strongly suggests that the 
observed color variations are real, and also indicates that for some asteroids 
color variations are stronger than for others.

\section{     The Apparently Random Nature of Color Variability  }

A series of tests discussed in the preceding section demonstrate that 
the detection of asteroid color variability is robust. In this section 
we attempt to find correlations between color variability and other 
parameters such as colors (a good proxy for taxonomic classes), absolute 
magnitude (i.e. size), and family membership.

%Fig. 11. 
\label{Fig11}
\begin{figure} 
\centering
\includegraphics[bb=20 20 501 822, width=\columnwidth]{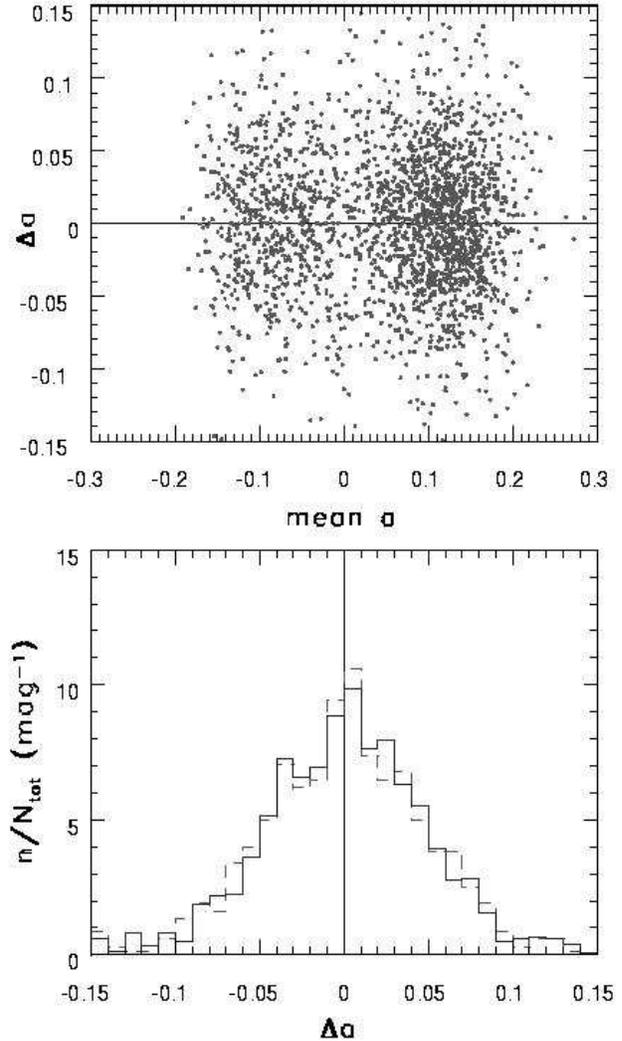} 
\caption{
The upper panel shows the $a$ color change between two epochs
for the same asteroids as in in Fig. 1, as a function of the mean $a$
color. The bottom panel compares the distributions of the $a$ color change 
for 689 objects with mean $a<0$ and 1585 objects with mean $a>0$. 
Note that the two histograms are statistically indistinguishable indicating 
that the color measurement is not correlated with the asteroid's $a$ color
(which is a good proxy for taxonomic classification).
}
\end{figure}

%Fig. 12. 
\label{Fig12}
\begin{figure} 
\centering
\includegraphics[bb=20 20 362 812, width=\columnwidth]{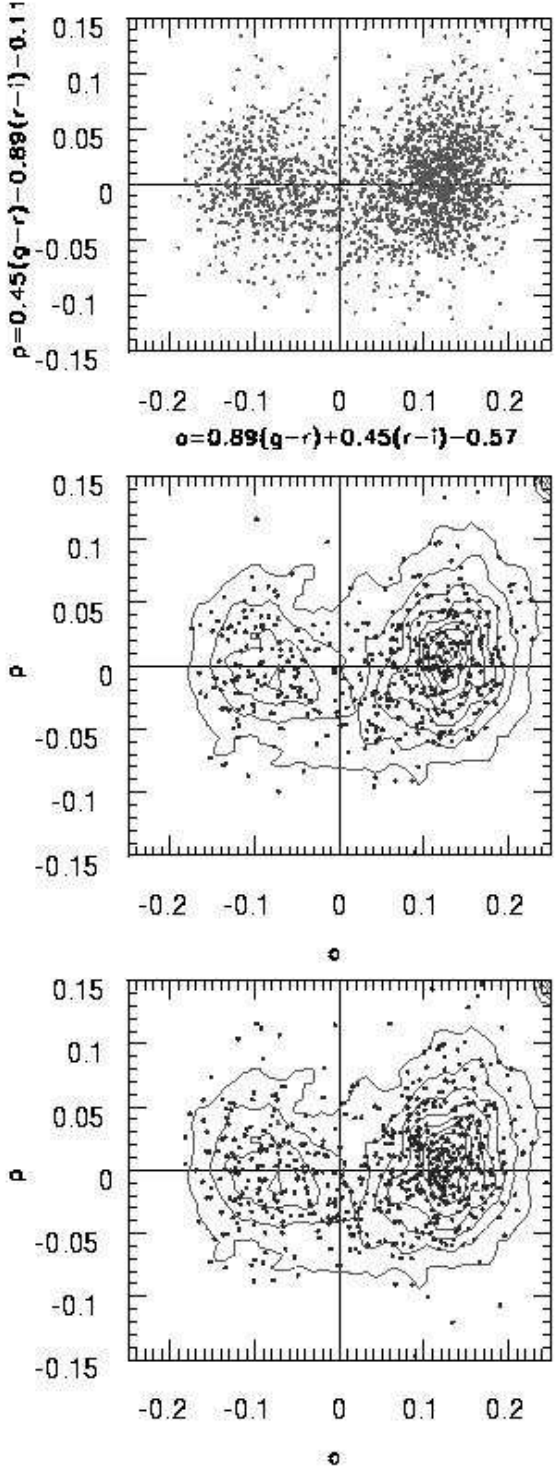} 
\caption{
The top panel shows the asteroid principal color diagram, constructed
with the mean colors for two measurements, for the whole sample. The
same distribution is shown by linearly spaced isodensity contours in
the middle and bottom panels. The dots in the middle panel represent
objects with the change in the $r$ magnitude of at least 0.2 mag. 
The dots in the bottom panel represent objects with the change in the 
$a$ color of at least 0.05 mag. Note that objects with large changes
of magnitudes and colors appear to show the same mean principal color 
distribution as the full sample.
}
\end{figure}

%Fig. 13. 
\label{Fig13}
\begin{figure} 
\centering
\includegraphics[bb=20 20 501 822, width=\columnwidth]{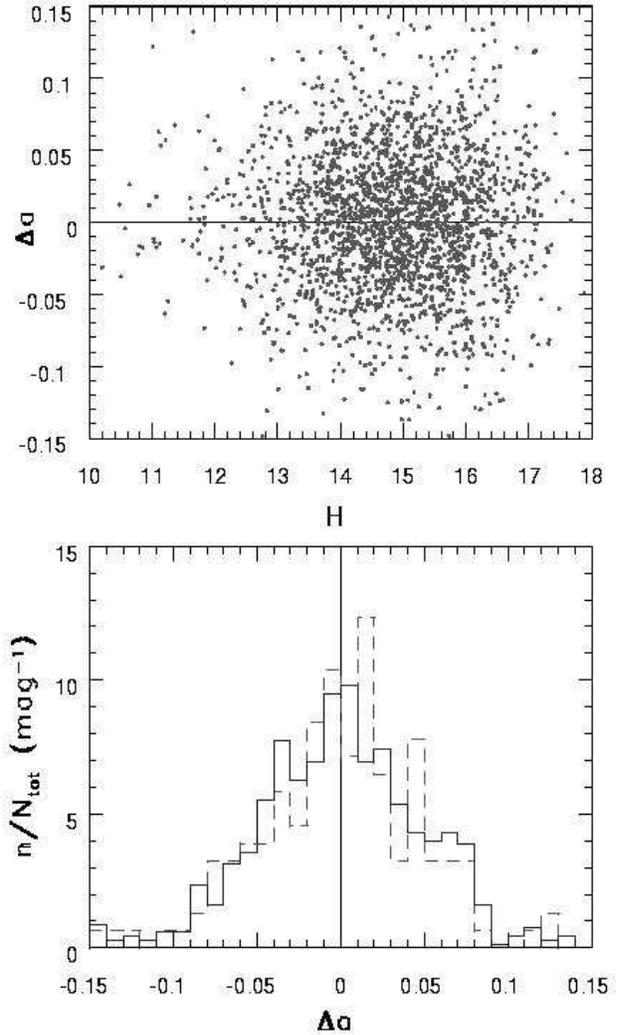} 
\caption{
The upper panel shows the $a$ color change between two epochs
for the same asteroids as in in Fig. 1, as a function of the absolute
magnitude. The bottom panel compares the width of the 
distribution of the $a$ color change for 97 objects with $10 < H < 13$ 
(dashed line) and for 505 objects with $15 < H < 16$ (solid line). 
Note that the two histograms are statistically indistinguishable indicating 
that the color measurement is not correlated with the asteroid's
absolute magnitude (i.e. size, in the approximate range 1--10 km).}
\end{figure}

\subsection{   Color Variability as a Function of Mean Colors }

The position of an asteroid in the principal color diagram (the top
panel in Fig. 8) is a good proxy for its taxonomic classification
(I01). Thus, a dependence of color variability on taxonomic
type would show up as a correlation between the change of $a$ color
and its mean value. We show the scatter plot of these two quantities
in the upper panel in Fig. 11. The bottom panel compares the distributions 
of the $a$ color change for 689 objects with mean $a<0$ (dominated by
the C type objects) and 1585 objects with mean $a>0$ (dominated by the
S type objects). The two histograms are statistically indistinguishable.

While the mean-color-selected objects (blue vs. red) appear to show 
the same $a$ color change distributions, it may be possible that the 
objects with the largest $\Delta r$ or $\Delta a$ would show different 
principal color distribution. We repeat the asteroid principal color 
diagram constructed with the mean colors (already shown in Fig. 8) 
in the top panel in Fig. 12. The same distribution is shown by linearly 
spaced isodensity contours in the middle and bottom panels. The dots in 
the middle panel represent objects with the change in the $r$ magnitude 
of at least 0.2 mag. The dots in the bottom panel represent objects with 
the change in the $a$ color of at least 0.05 mag. There is no discernible
difference between the color distribution for the whole sample and that
for the highly variable objects.

\subsection{ Color Variability as a Function of Absolute Magnitude  }

Asteroids of different size may exhibit different color variability.
The closest proxy for the size, in the absence of direct albedo measurements, 
is the absolute magnitude. Fig. 12 shows correlation between $\Delta a$ and the 
absolute magnitude $H$. The displayed range of $H$ roughly corresponds to 
the 1--10 km size range. The bottom panel compares the width of the $\Delta a$
distribution for 97 objects with $10 < H < 13$ (dashed line) and for 505 
objects with $15 < H < 16$ (solid line). The two histograms are statistically 
indistinguishable indicating that the color change is not correlated with the 
asteroid's absolute magnitude (i.e. size). However, we emphasize that the dynamic
range of probed sizes is fairly small.

\subsection{Color Variability as a Function of Family Membership}

SDSS colors are a good proxy for taxonomic classification, and can be 
efficiently used to recognize at least three color groups (J02). I02c 
showed by correlating asteroid dynamical families and SDSS colors that 
indeed there are more than just three ``shades'': many families have
distinctive and uniform colors. Motivated by their finding, we obtain
a more detailed classification of asteroids using dynamical clustering
and correlate it with variability properties. 

We define families by 3-dimensional boxes in the space spanned by proper
orbital elements (the boundaries are summarized in Table 1), using the results 
from Zappala et al. (1995). There are 6 families with more than 50 members 
in the sample analyzed here. The comparison of $|\Delta a|$ and $|\Delta r|$ 
distributions for individual families with those for the whole sample are shown 
in Figs. 14 and 15, respectively. In order to quantitatively assess whether 
there is any family that differs in its variability properties from the rest 
of the sample, we performed two-sample Kolmogorov-Smirnov tests. None of
the families listed in Table 1 was found to differ from the mean values
for the whole sample at a confidence level greater than 95\%. We conclude that 
all the examined families show similar variability properties.

\begin{table}
\label{Table1}
\caption{The definitions of asteroid families in the $a_p$-$sin i$-$e$ space}
\begin{tabular}{llll}
\hline
Groups & $a_p$ & $\sin(i)$ & $e$\\
\hline
Flora & 2.16--2.32 & 0.04--0.125 & 0.105--0.18\\
Vesta & 2.28--2.41 & 0.10--0.135 & 0.07--0.125\\
Nysa-Polana & 2.305--2-48 & 0.03--0.06 & 0.13--0.21\\
Eunomia-Adeona & 2.52--2.72 & 0.19--0.26 & 0.12--0.19\\
Eos & 2.95--3.10 & 0.15--0.20 & 0.04--0.11\\
Themis & 3.03--3.23 & 0--0.6 & 0.11--0.20\\ 
\hline
\end{tabular}
\end{table}

%Fig. 14 
\label{Fig14}
\begin{figure*} 
\centering
\includegraphics[bb=50 150 302 300, width=5.5cm]{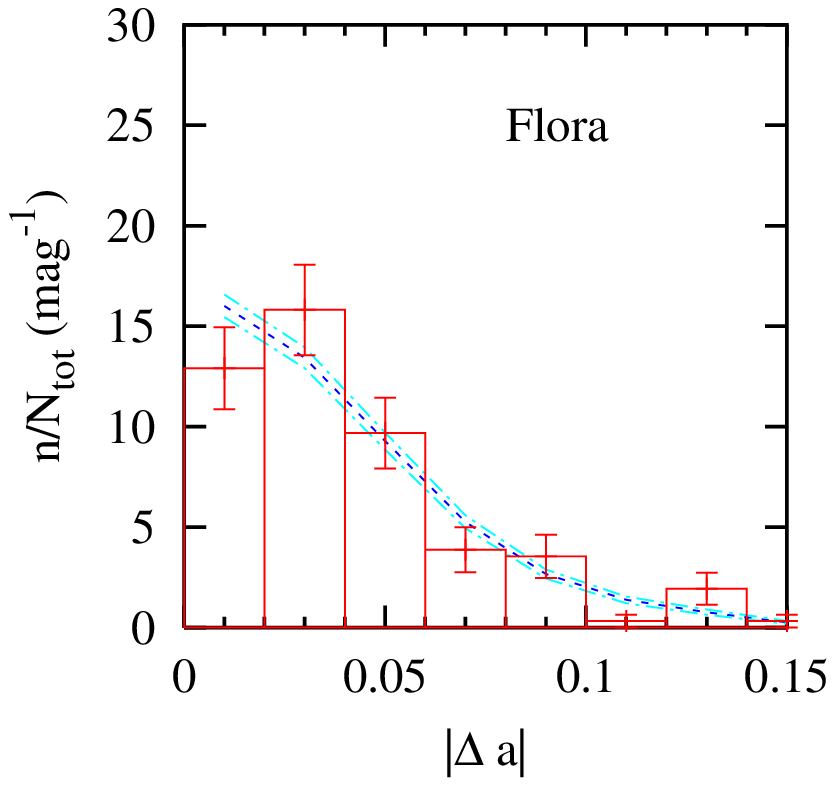} 
\includegraphics[bb=50 150 302 300, width=5.5cm]{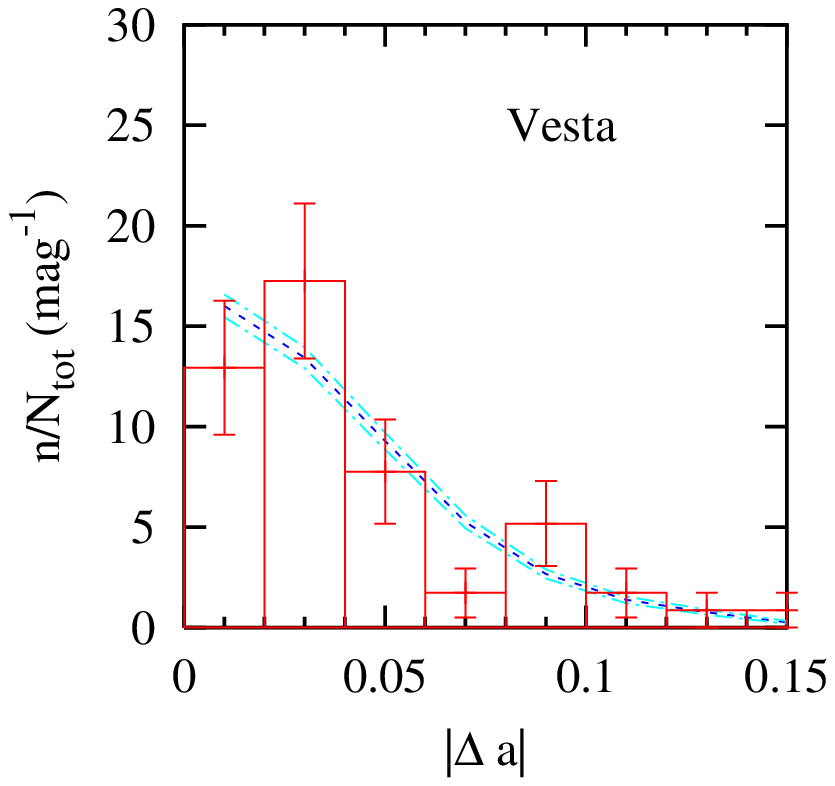} 
\includegraphics[bb=50 150 302 300, width=5.5cm]{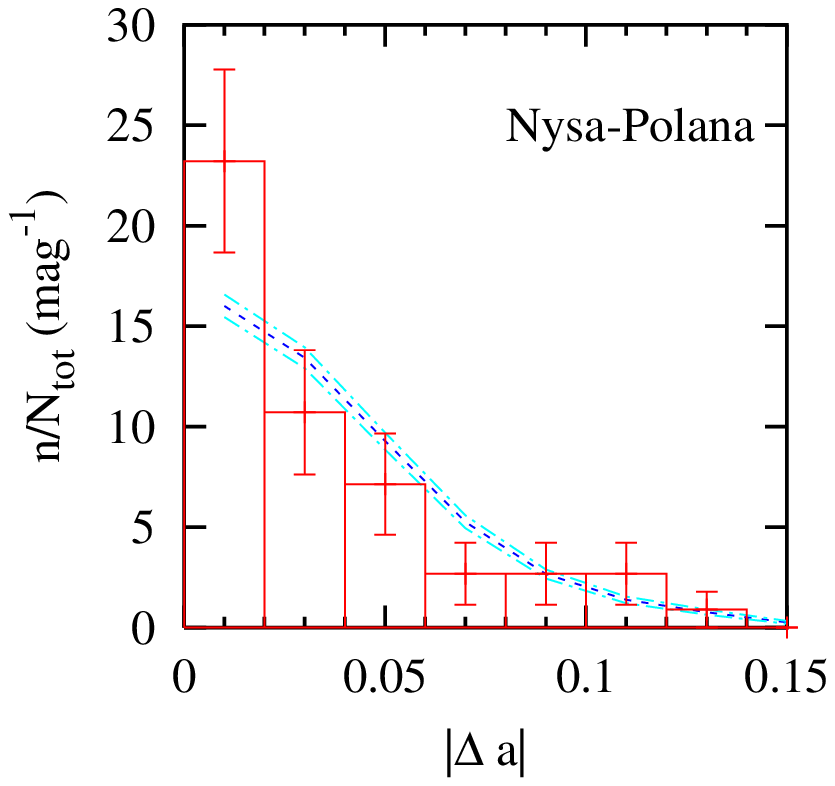} 
\includegraphics[bb=50 50 302 380, width=5.5cm]{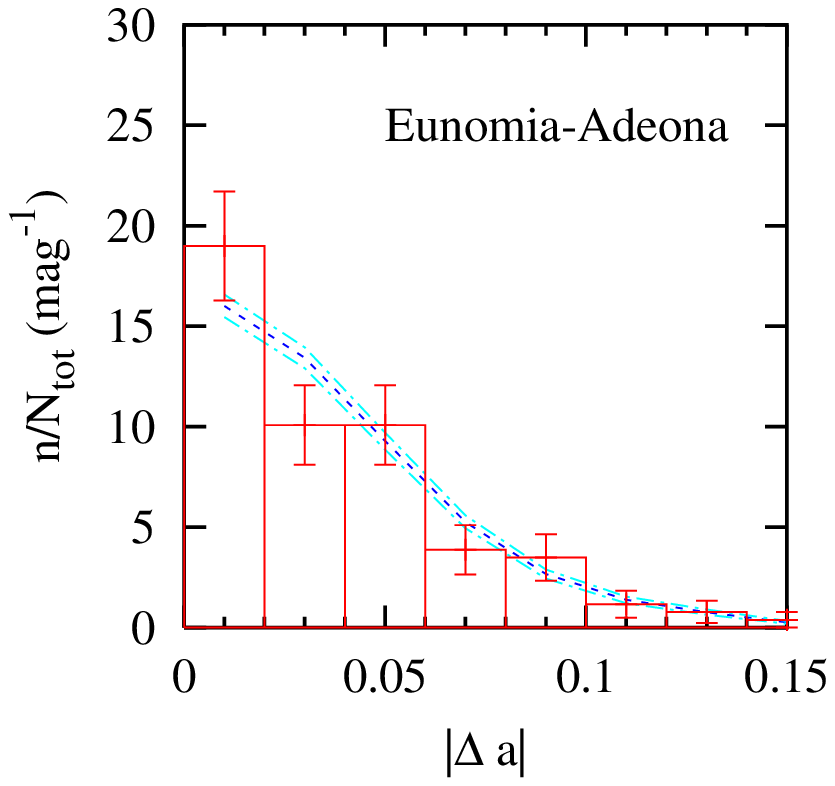} 
\includegraphics[bb=50 50 302 380, width=5.5cm]{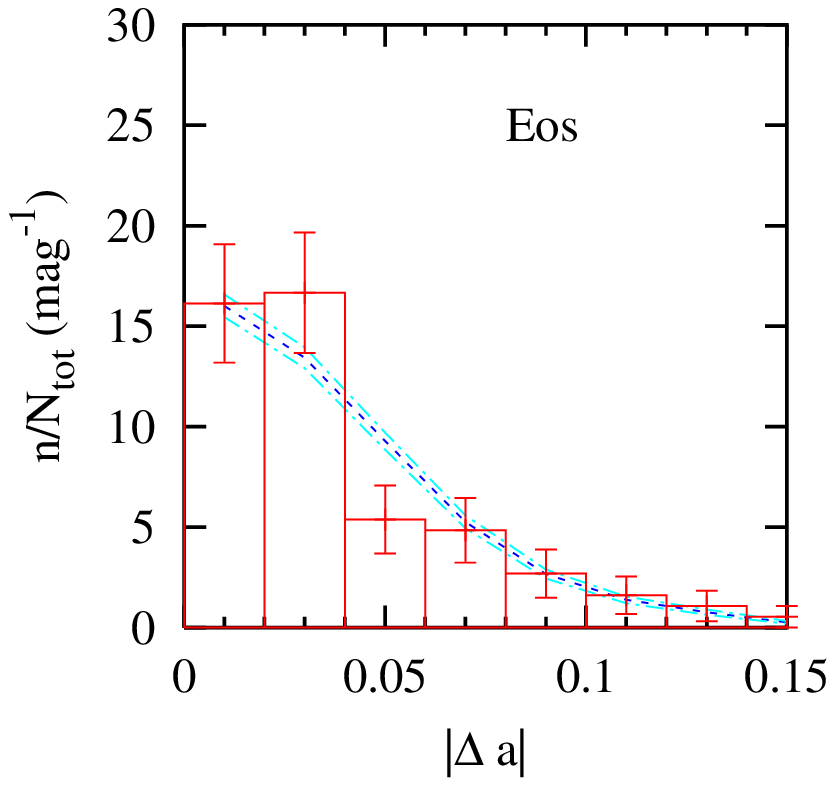} 
\includegraphics[bb=50 50 302 380, width=5.5cm]{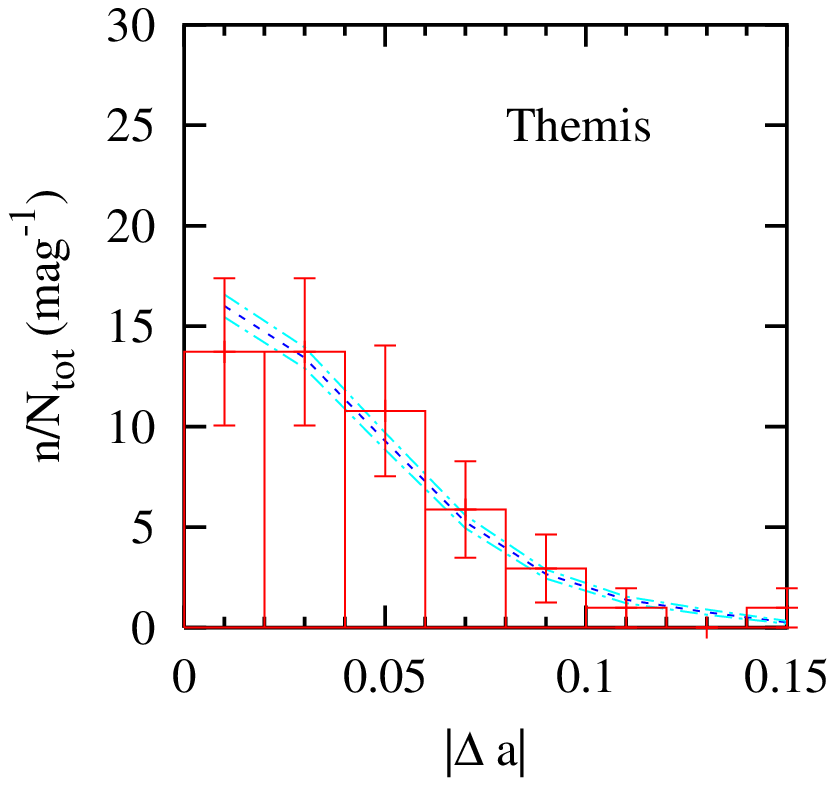} 
\caption{The distribution of $|\Delta a|$ for individual asteroid families 
(boxes with errorbars) is compared to the distribution for the whole sample,
shown by lines (mean $\pm\sigma$). The families are defined by regions in 
the spaced spanned by proper orbital elements (see Table 1. and text).
}
\end{figure*}

%Fig. 15
\label{Fig15}
\begin{figure*} 
\centering
\includegraphics[bb=50 150 302 300, width=5.5cm]{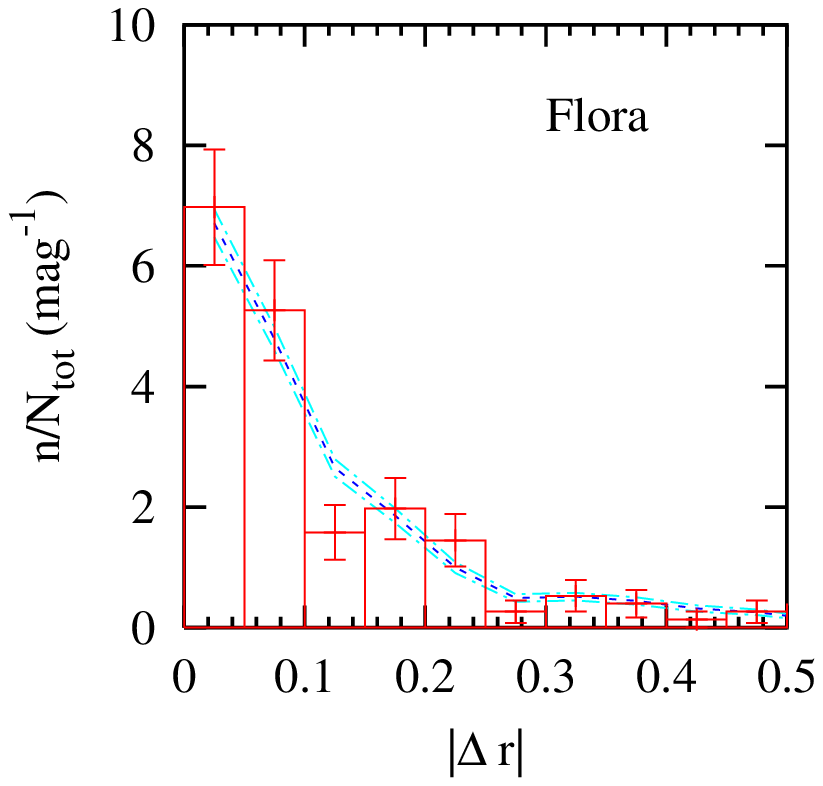} 
\includegraphics[bb=50 150 302 300, width=5.5cm]{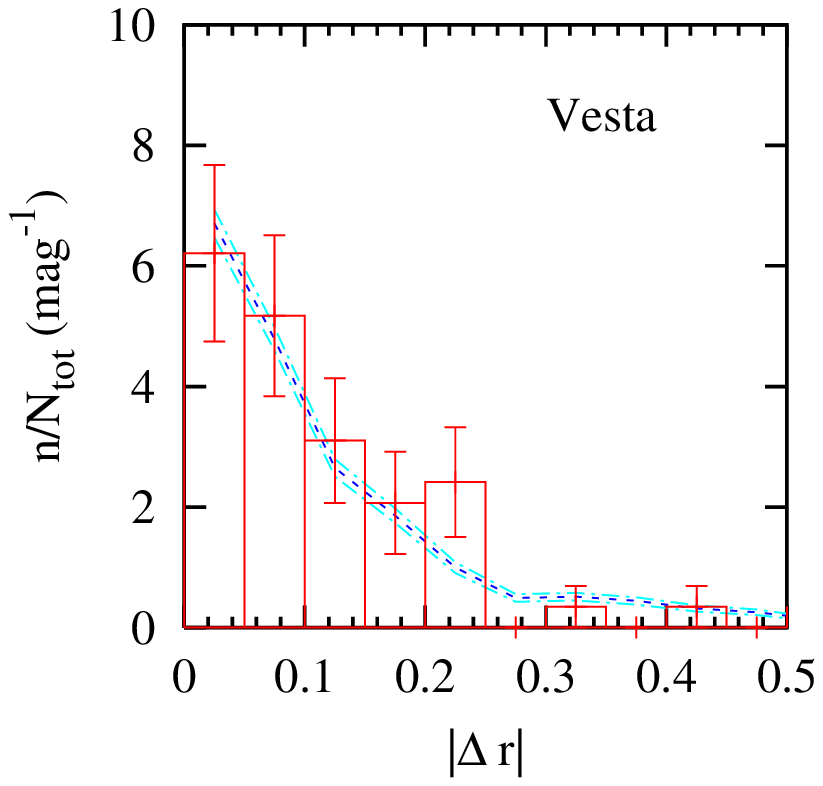} 
\includegraphics[bb=50 150 302 300, width=5.5cm]{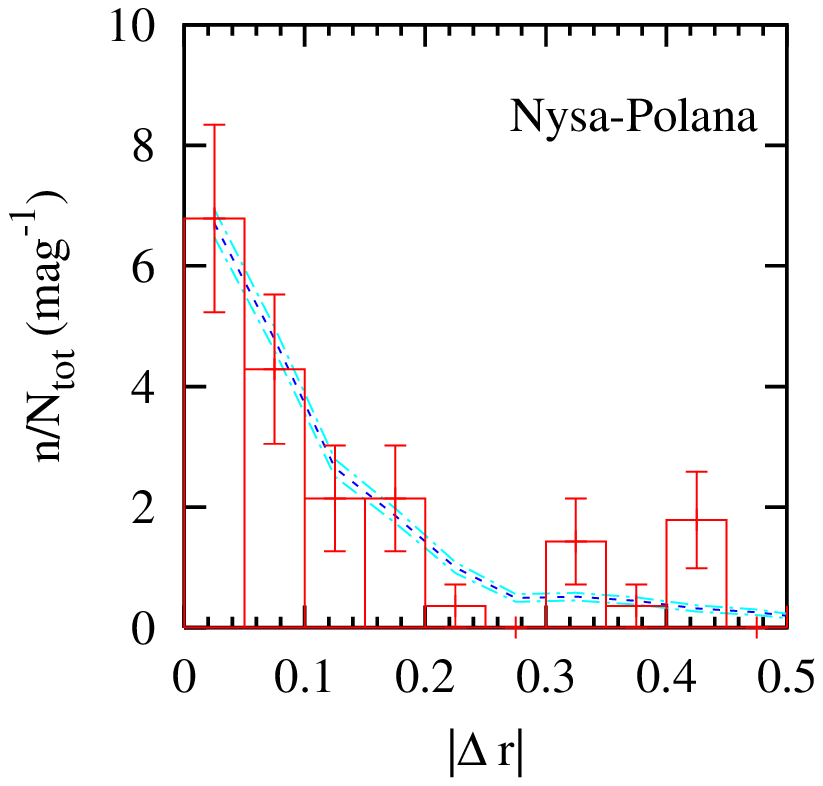} 
\includegraphics[bb=50 50 302 380, width=5.5cm]{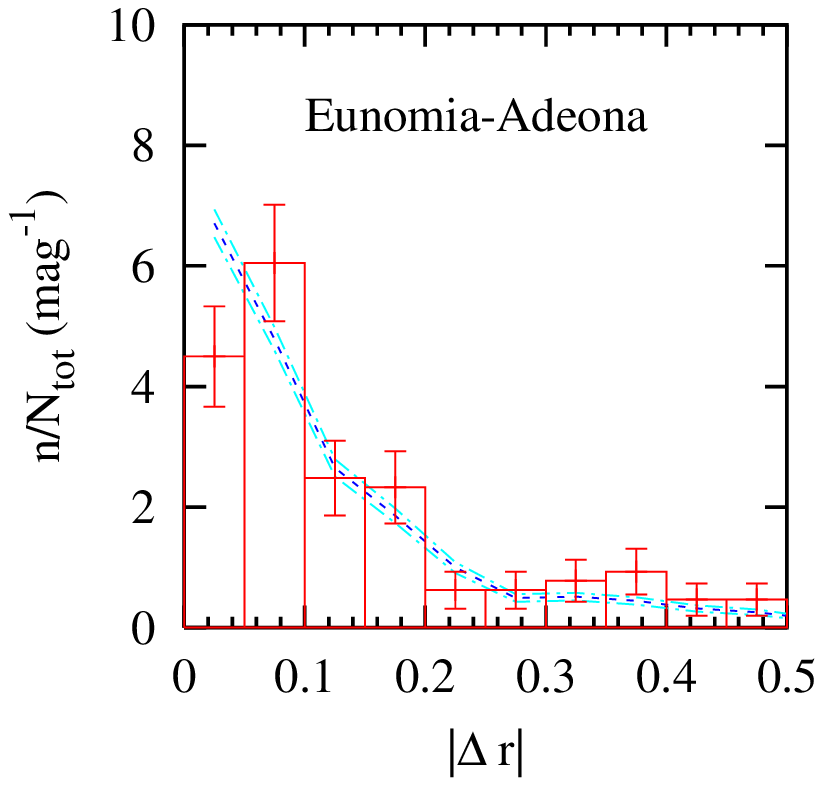} 
\includegraphics[bb=50 50 302 380, width=5.5cm]{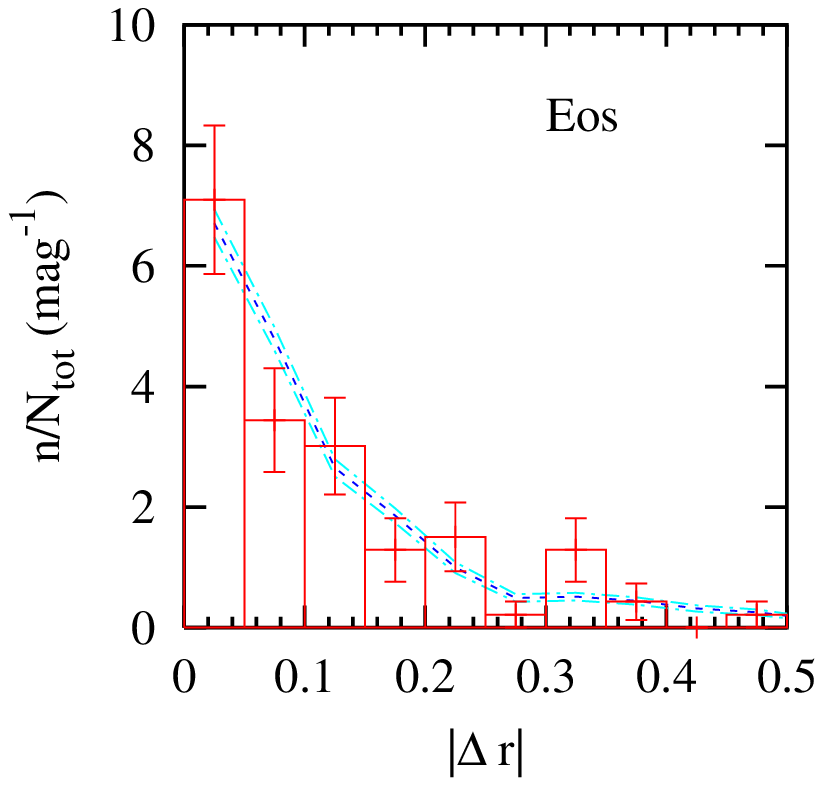} 
\includegraphics[bb=50 50 302 380, width=5.5cm]{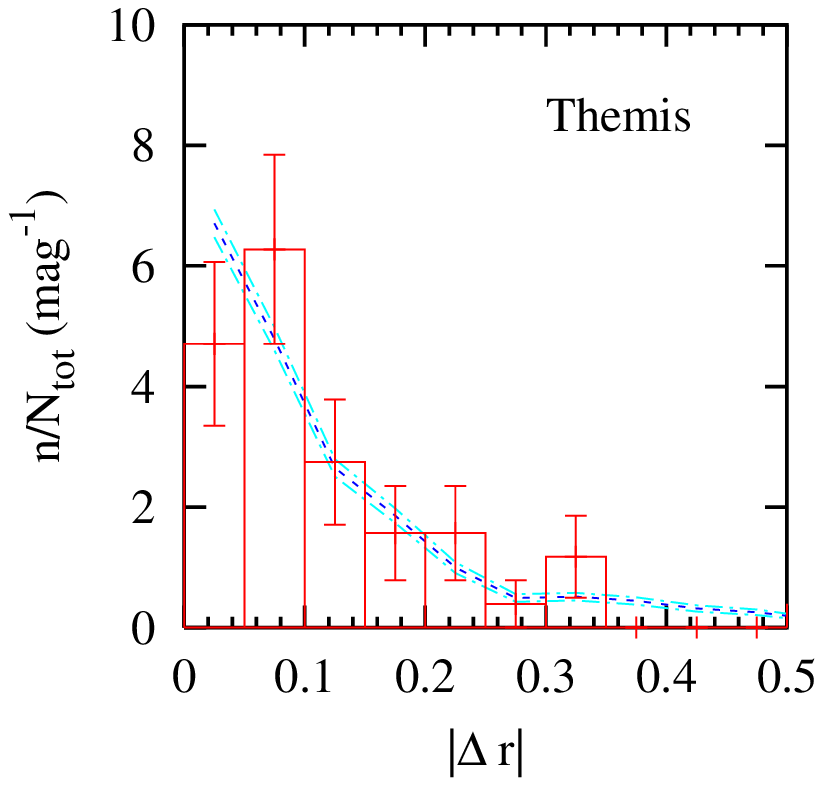} 
\caption{Same as Figure 14, except that the absolute value of the
$r$ band brightness variation is shown.
}
\end{figure*}

\section{Discussion and Conclusions}

The detection of color variability for a large sample of asteroids discussed 
here represents a significant new constraint on the physical properties 
and evolution of these bodies. The random nature of this 
variability, implied by the lack of apparent correlations with asteroid 
properties such as mean colors, absolute magnitude and family membership,
could be interpreted as due to some hidden random photometric error. However, 
several lines of evidence argue that this explanation is unlikely. First,
the magnitude of the observed effect (0.06--0.11 mag.) is so large that 
such photometric errors would have to be noticed in numerous other studies and
tests based on SDSS photometric data. If such an error only shows up for 
moving objects, then the color variability should increase with the apparent 
velocity, an effect which is not observed (see Fig. 2). Second, independent pairs 
of observations, discussed in Section 3.4, seem to ``know'' about each other: 
objects with large color variation in one pair of observations tend to show 
relatively large color variation in other pair of observations. 
This fact cannot be explained by random photometric errors. Third, the color 
variation is not entirely random in the principal colors diagram, as discussed in 
Section 3.3. 
There is a preferred direction for variability-induced motion in this diagram, 
and the scatter in principal colors is significantly different ($\sigma_p \sim 
1.33 \sigma_a$). Were the color variability caused by random photometric errors, 
it would not be correlated with the distribution of asteroid principal colors.  

The observed color variability implies inhomogeneous albedo distribution over 
an asteroid surface. Although the color variability is fairly small, it 
suggests that large patches with different color than their surroundings exist 
on a significant fraction of asteroids. For example, consider a limiting case of
an asteroid with two different hemispheres, one with C type material, and one 
with S type material. In this case the peak-to-peak amplitude of its $a$ color 
variability would be only 0.2 mag., with rms$\sim$0.05 mag.\footnote{The color 
lightcurve would be biased red because of the factor of 4 difference in the visual 
albedos.}, even under the most favorable condition of the rotational axis 
perpendicular to the line of sight. Taking into account a distribution of the 
angle between rotational axis and the line of sight would decrease these values 
further. Without detailed modeling it is hard to place a lower limit on the 
fraction of surface with complementary color to explain the observed color 
variations. However, using simple toy models and colors typical for C and 
S type asteroids, we find that this fraction must be well over 10\%.
The features seen in spatially resolved color images of Eros obtained by NEAR 
spacecraft (Murchie et al. 2002) support such a conclusion. 

A simple explanation for the existence of patches differing in color from their 
surroundings is the deposition of material (e.g. silicates on C type asteroids 
and carbonaceous material on S type asteroids) by asteroid collisions. However, 
such surfaces would exhibit color variations preferentially aligned with 
the $a$ color axis, contrary to the observations. For a given fraction of 
asteroid surface affected by the deposition of new material, the large difference
in albedos of silicate and carbonaceous surfaces would probably produce 
different amplitudes of color variability for S and C type asteroids (due 
to large difference in their albedos), a behavior that is not supported by the 
data. Thus, the color variability cannot be explained as due to patches of
S-like and C-like material scattered across an asteroid surface.
 
A plausible cause for optically inhomogeneous surface is space weathering. 
This phenomenon includes the effects of bombardment by micrometeorids, cosmic rays, 
solar wind and UV radiation, and may alter the chemistry of the surface material
(Zeller \& Rouca, 1967). Recent spacecraft data indicate that these processes 
may be very effective in the reddening and darkening of asteroid surface 
(Chapman 1996). In this interpretation the $u-g$ color should show the largest 
variation (Hendrix \& Vilas 2003), and this is indeed supported by the data 
presented here (see Fig. 7). It is not clear, however, how could such processes 
result in fairly large isolated surface inhomogeneities.

An interesting explanation for surface inhomogeneities is the effects of
cratering. Using NEAR spacecraft measurements, Clark et al. (2001) find 
30-40\% albedo variations on the Psyche crater wall on Eros. Such a large
effect could perhaps produce color variations consistent with observations.
However, it seems premature to draw conclusions with detailed modeling.  

Irrespective of the mechanism(s) responsible for the observed color variations, 
our results indicate that this is a rather common phenomenon. We conclude by 
pointing out that the sample presented here will be enlarged by a factor of 
few in a year or two because SDSS is still collecting data, and the size of 
known object catalog, needed to link the observations, is also growing. 
Furthermore, the faint limit of the known object catalog is also improving, 
and will result in a larger size range probed by the sample. Apart from 
SDSSMOC, the upcoming (in 5--10 years) deep synoptic surveys such as Pan-STARRs 
and LSST will produce samples of size and quality that will dwarf the sample 
discussed here, and provide additional clues about the causes of asteroid color
variability.

\section*{Acknowledgments}

This work has been supported by the Hungarian OTKA Grants
T034615, FKFP Grant 0010/2001, Szeged Observatory Foundation, and 
Pro Renovanda Cultura Hungariae Foundation DT 2002/maj.21. 
\v{Z}.I. acknowledges generous support by Princeton University.

Funding for the creation and distribution of the SDSS Archive has been
proved by the Alfred P. Sloan Foundation, the Participating Institutions,
the National Aeronautics and Space Administration, the National Science
Foundation, the U.S. Department of Energy, the Japanese Monbuhagakusho, and
the Max Planck Society. The SDSS Web site is www.sdss.org. The Participating
Institutions are The University of Chicago, Fermilab, the Institute for
Advanced Study, the Japan Participation Group, The Johns Hopkins University,
the Max-Planck-Institute for Astronomy (MPIA), the Max-Planck-Institute for
Astrophysics (MPA), New Mexico State University, Princeton University, the
United States Naval Observatory, and the University of Washington.


\begin{thebibliography}{}

\bibitem[]{} Azebajian, K. et al., 2003, AJ, in press

\bibitem[]{} Blanco,C., Catalano, S. 1979, Icarus, 40, 359
  
\bibitem[]{} Binzel, R.P., et al., 1997, Icarus, 128, 95

\bibitem[Bowell \& Lumme 1979]{BL79} Bowell, E. \& Lumme, K., 1979, in Asteroids,
         ed. T. Gehrels, (Tuscon: Univ. of Arizona Press), 132

\bibitem[]{AstorbURL} Bowell, E. 2001, {\it Introduction to ASTORB}, available from 
        ftp://ftp.lowell.edu/pub/elgb/astorb.html

\bibitem[]{} Chapman, C.R. 1996, Meteoritics, 31, 699

\bibitem[]{} Clark, B.E. et al. 2001, Meteoritics and Planetary
             Science, 36, 1617 

\bibitem[]{} Degewij, J., Tedesco, E.F., Zellner, B.H. 1979, Icarus, 40, 364

\bibitem[Efron \& Petrosian 1992]{EP92} Efron, B. \& Petrosian, V. 1992, ApJ,
         399, 345

\bibitem[]{fukugita} Fukugita, M., et al. 1996, AJ, 111, 1748

\bibitem[]{gunn} Gunn, J.E., et al. 1998, AJ, 116, 3040

\bibitem[]{} Hendrix, A.R., Vilas, F. 2003, 34th Annual Lunar and Planetary
     Science Conference, March 17-21, 2003, League City, Texas
 
\bibitem []{} Ivezi\'{c}, \v{Z}., Tabachnik, S., Rafikov, R., et al. 2001, AJ, 122, 2749 (I01)

\bibitem[]{} Ivezi\'c, \v{Z}., Juri\'c, M., Lupton, R.H., et al. 
        2002a, Survey and Other Telescope Technologies and Discoveries, 
        J.A. Tyson, S. Wolff, Editors, Proceedings of SPIE Vol. 4836 (2002),
        also astro-ph/0208099

\bibitem[]{} Ivezi\'c, \v{Z}., Lupton, R.H., Juri\'c, M., et al.
        2002b, AJ 124, 2943

\bibitem[]{} Ivezi\'c, \v{Z}., Lupton, R.H., Anderson, S., et al. 
         2002c, astro-ph/0301400

\bibitem[]{}
   Gaffey, M.J., King, T., Hawke, B.R., 1982, Workshop on Lunar Breccias and
   their Meteoritic Analogs, ed. by G.J. Taylor and L.L. Wilkening, LPI, Huston

\bibitem[]{} Juri\'{c}, M., Ivezi\'{c}, \v{Z}, Lupton, R.H., et al., 2002, Astronomical Journal, 124, 1776

\bibitem[]{lupton} Lupton, R.H., Gunn, J.E., Ivezi\'{c}, \v{Z}., et al., 2001, 
       in {\it Astronomical Data Analysis Software and
       Systems X}, ASP Conference Proceedings, Vol.238, p. 269.
       Edited by F. R. Harnden, Jr., Francis A. Primini, and Harry E. Payne. San Francisco:
       Astronomical Society of the Pacific, ISSN: 1080-7926

\bibitem[]{} Lupton, R.H., 1993, Statistics in theory and practice, 
             Princeton, N.J.: Princeton University Press

\bibitem[]{} Magnusson, M., 1991, A\&A, 243, 512

\bibitem[]{} Melillo, F.J., 1995, MPBu 22, 42

\bibitem[]{} Metcalf, J.H., 1907, ApJ, 25, 264

\bibitem[]{} Milani, A. et al., 1999, Icarus, 137, 269

\bibitem[]{} Mottola, S., Gonano-Beurer, M., Green, S.F., et al., 1994, P\&SS, 44, 21

\bibitem[]{} Murchie, S., Robinson, M., Clark, B., et al., 2002, Icarus, 155, 145

\bibitem[]{} Pravec, P., Harris, A.W., 2000, Icarus, 148, 12

\bibitem[]{} Reed, K.L., Gaffey, M.J., Lebofsky, L.A., 1997, Icarus, 125, 446

\bibitem[]{} Schober, H.J., Schroll, A., 1992, in: Sun and Planetary System,
   Dordrecht, D. Reidel Publishing Co., 1982, 258

\bibitem[]{} Sekiguchi, T., Boehnhardt, H., Hainaut, O. R., Delahodde, C. E., 2002,
   A\&A, 385, 281

\bibitem[]{} Sheppard, S.S., Jewitt, D.C., 2002, AJ, 124, 1775

\bibitem[Shoemaker {\em et al.} 1979]{Setal79} Shoemaker, E., Williams, J.G.,
         Helin, E.F., \& Wolfe, R.F. 1979, in Asteroids,
         ed. T. Gehrels, (Tuscon: Univ. of Arizona Press), 253

\bibitem[]{smith} Smith, J.A. {\it et al.}, 2002, AJ, 123, 2121

\bibitem[]{EDR} Stoughton, C. {\it et al.}, 2002, AJ, 123, 485

\bibitem[]{} Szab\'o, Gy.M., Kiss, L.L., S\'arneczky, K., et al., 2001, A\&A 384, 702

\bibitem[]{} Thomas, P. C., Binzel, R. P., Gaffey, M. J., et al.,
              1997, Science, 277, 1492

\bibitem[]{} Zappala, V, Bemdjoya, Ph., Cellino, A., Farinella, P., Froeschl\'e, C.
             1995, Icarus, 116, 291

\bibitem[]{} Zeller, E.J., Rouce, L.B., 1967, Icarus, 7, 372

\bibitem[Zellner 1979]{Zellner79} Zellner, B. 1979, in Asteroids, ed. T. Gehrels,
         (Tuscon: Univ. of Arizona Press), 783

\bibitem[]{Zellner85} Zellner, B., Tholen, D.J. \& Tedesco, E.F., 1985, Icarus, 61, 355

\end{thebibliography}
\end{document}